\input epsf
\input lecproc.cmm

\topinsert\vbox{\hrule\smallskip\hbox{\quad Procs. Conference
`Relativistic Jets in AGNs', Cracow, May 1997\quad}\smallskip\hrule}\endinsert

\contribution{X-ray Emission from Jets in Centrally Obscured AGN}

\author{D.M. Worrall}

\address{Department of Physics, University of Bristol, U.K. \&
Harvard-Smithsonian Center for Astrophysics, U.S.A.}

\vskip -\baselineskip

\abstract{There is good evidence for X-ray emission associated
with AGN jets which are relativistically boosted towards the observer.
But to what jet radius does such X-ray emission persist?  To attempt
to answer this question one can look at radio galaxies; their cores
are sufficiently X-ray faint that any unbeamed X-ray emission in the
vicinity of the central engine must be obscured.  The jets of such
sources are at unfavourable angles for relativistic boosting, and so
their relatively weak X-ray emission must be carefully separated from
the plateau of resolved X-ray emission from a hot interstellar,
intragroup, or intracluster medium on which they are expected to sit.
This paper presents results arguing that jet X-ray emission is
generally detected in radio galaxies, even those of low intrinsic
power without hot spots.  The levels of emission suggest an
extrapolated radio to soft X-ray spectral index, $\alpha_{rx}$, of
about 0.85 at parsec to perhaps kiloparsec distances from the cores.
}

\vskip -2\baselineskip

\titlea{1}{High-Power Jets}
Other papers in this volume amply illustrate an association of the
X-ray emission of $\gamma$-ray emitting quasars and BL Lac objects
with synchrotron emission or Compton scattering in a relativistic jet
strongly boosted in the line of sight towards the observer. However,
the correlation of X-ray emission with core-radio emission which is so
striking in core-dominated quasars (see e.g. Fig.~1) does not extend
to lobe-dominated quasars, suggesting that in addition to highly
anisotropic jet-related emission there is an X-ray component which is
considerably more isotropic (Zamorani 1984; Worrall et al.~1987).
X-ray spectral measurements lend support to the idea that this second
component (1) is associated with emission from the vicinity of the
central engine and (2) may be similar to the dominant X-ray component
in radio-quiet quasars (Wilkes and Elvis 1987).

\begfig 0cm
\vbox{
\vskip -\baselineskip
\centerline{\epsfxsize=.5\hsize
\epsfbox{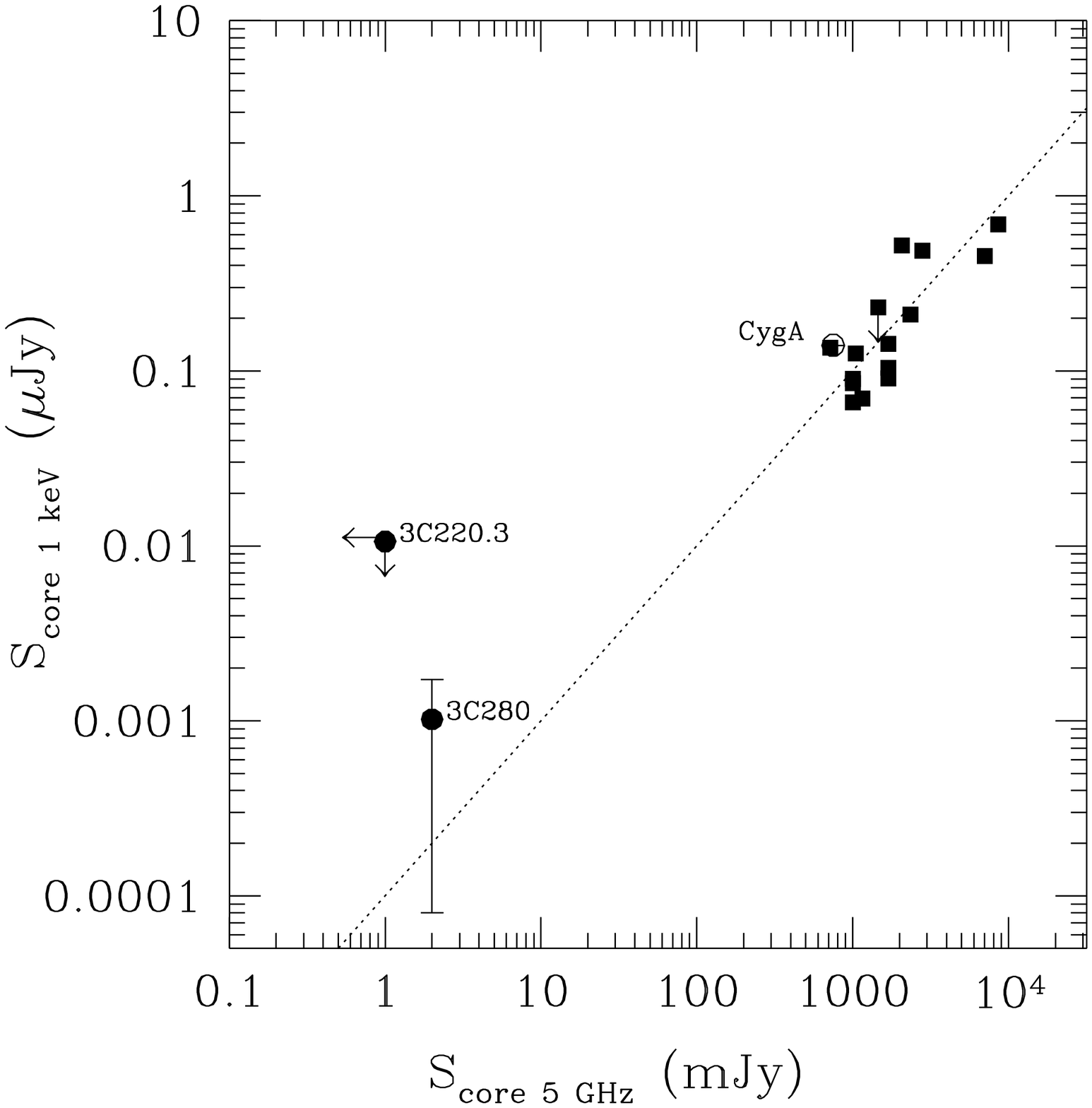}
 \hfil
\epsfxsize=.5\hsize
 \epsfbox{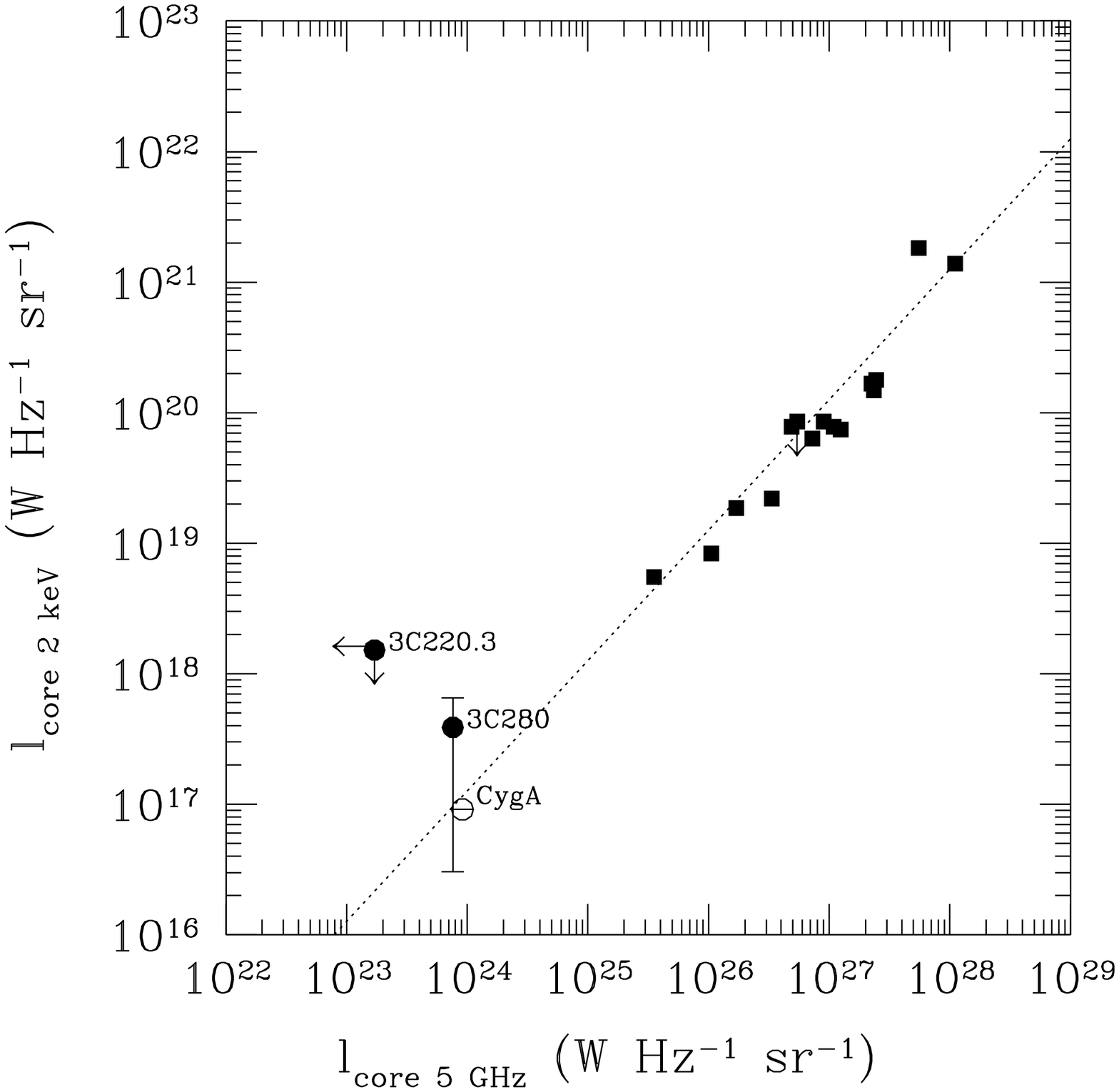} }
\vskip -\baselineskip
{\figure{1}{
3C quasars and radio galaxies matched in isotropic radio power and
redshift: Total ($=$ core) X-ray {\it vs\ } core radio emission
for core-dominated
quasars (squares) and core X-ray {\it vs\ } core radio for radio
galaxies 3C~280 and 3C~220.3 (circles) from Worrall et al.~(1994).
See text for discussion of Cygnus~A.  Flux-density plot on the left;
Luminosity density on the right [$H_o = 50 $~km s$^{-1}$
Mpc$^{-1}$]. Lines of slope unity are drawn for reference, and the
correlation gives an extrapolated radio to X-ray spectral index
$\alpha_{rx} \simeq 0.86$.
}}
}
\endfig

The X-ray central component is either absent or obscured in radio galaxies,
since as a class they are considerably fainter than radio-quiet
quasars.  Obscuration is in line with Unification models, and implies
that any AGN-related X-ray emission measured in radio galaxies is
either seen through a large absorbing column of neutral gas or
originates in the jet at a radius beyond the obscuration.

The 3C radio survey selects the sources brightest in low-frequency
(and thus predominantly unbeamed) radio emission, but the
distributions of total unbeamed radio power (a key parameter in the
selection of sources believed to differ only though orientation) for
its quasars and galaxies overlap only at the highest powers (Laing,
Riley \& Longair 1983).  In a flux-limited survey like 3C, high power
means a bias towards high redshift. So, to view a powerful radio-loud
quasar oriented as a twin-jet radio source we are forced towards
redshifts $\geq 0.4$, where only very long exposures with ROSAT detect
radio galaxies in the X-ray (e.g., Crawford \& Fabian 1996).  Where
enough counts are present to measure source extent, it is generally
found that the galaxies are resolved, and so much or all of the
emission is thought to be associated with X-ray emitting gas of
cluster dimension.

For the $z = 1$ radio galaxy 3C~280, Worrall et al.~(1994) were so bold
as to attempt a separation of resolved and unresolved X-ray emission.
They found that the unresolved X-rays fitted on an extrapolation of
the X-ray to core-radio proportionality found for core-dominated
quasars (Fig.~1), and claimed this as evidence for a jet related
component of X-ray emission in the host radio galaxies of quasars.

Cygnus~A, with a 5~GHz $0.4''$-resolution core-radio flux density of
750~mJy (Carilli \& Barthel~1996) is well matched to 3C~280 in core
and total radio power and is unique for its proximity.  X-ray spectra
taken with EXOSAT and {\it Ginga\ } have found evidence for core
emission absorbed by a column density of $\sim 4 \times
10^{23}$~cm$^{-2}$ (Arnaud et al.~1987; Ueno et al.~1994). The high
absorption implies this component should not have been detected in the
42~ks ROSAT HRI exposure because of the HRI's lower-frequency X-ray 
response, and yet an unresolved core X-ray component of $270 \pm 60$
net counts was seen (Harris et al.~1994).  I suggest that the
low-frequency HRI core component instead arises from the jet in
regions where the only absorption along the line of sight is the
Galactic value of $3.3
\times 10^{21}$ cm$^{-2}$.  For a power-law spectrum of energy index
1.0 the HRI counts correspond to a 1~keV flux density of
0.13~$\mu$Jy.  With this assumption, I have plotted Cygnus~A on
Fig.~1.  The remarkable agreement with the correlation suggests that
the correct interpretation for the low-energy X-ray core of Cygnus~A
is unobscured jet-related emission.

The search for jet-related X-ray emission from powerful radio galaxies
must concentrate on soft X-ray energies, where the contribution
from highly absorbed core X-ray emission is negligible, and
galaxies devoid of the broad emission-lines which may
indicate a partially unobscured view of the central engine.
Results are encouraging and suggest that the X-ray emission from
powerful radio galaxies should not be interpreted merely as a mixture
of absorbed-core and hot-gas components.

\titlea{2}{Low-Power Jets}
Strong cases have been made that BL~Lac objects and quasars should be
treated separately because of divergent properties and likely
different host populations: low-power FRI radio galaxies for BL~Lacs,
and high-power, hotspot, FRII radio galaxies for quasars (e.g., Browne
1989).  I've argued that for high-power sources the viewing-angle
progression goes from radio galaxy (where X-rays from the central
engine suffer obscuration) to lobe-dominated quasar (where X-rays from
the central engine can be seen) to core-dominated quasar (where X-rays
from the central engine become swamped by beamed jet emission).
However, the equivalent of lobe-dominated BL Lac objects are not
readily identified.  This begs the question as to whether or not the
central engine of a BL Lac object is X-ray bright.  What is clear is
that unresolved X-ray emission from low-power radio galaxies is
sufficiently faint and well correlated with core radio strength (see
later) that as in the case of quasar host galaxies, any substantial
unbeamed X-ray emission in the vicinity of the central engine must be
obscured.

Ulrich (1989) has shown that the B2 radio-galaxy sample (complete set
of 50 sources from the B2 radio survey, $S_{\rm 408 MHz} \geq 0.25$ Jy
in 6.7\% of the sky, identified with elliptical galaxies brighter than
$m_{\rm ph} = 15.7$) is well matched to radio-selected BL~Lac objects
in their extended radio properties and galaxy magnitudes.  This
sample, now mostly observed in ROSAT pointed observations with the
PSPC or the HRI or both, is therefore an ideal testbed for Unification
models and derivation of X-ray beaming factors using methods similar
to those of Padovani \& Urry (1990), but with the added advantage of
X-ray component separation.

\begfig 0cm
\vbox{
\vskip -\baselineskip
\centerline{\epsfxsize=.5\hsize
\epsfbox{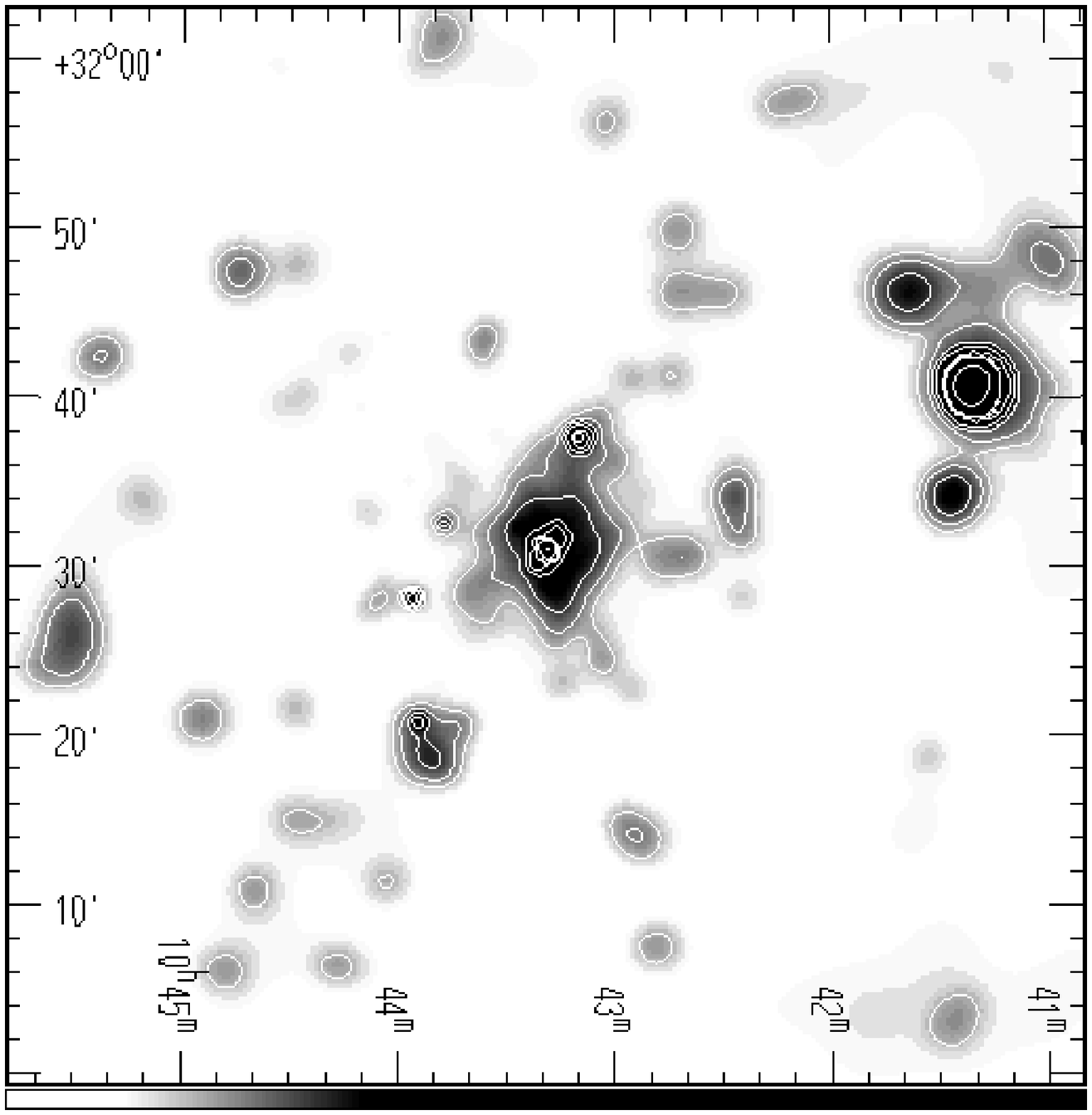}
 \hfil
\epsfxsize=.5\hsize
 \epsfbox{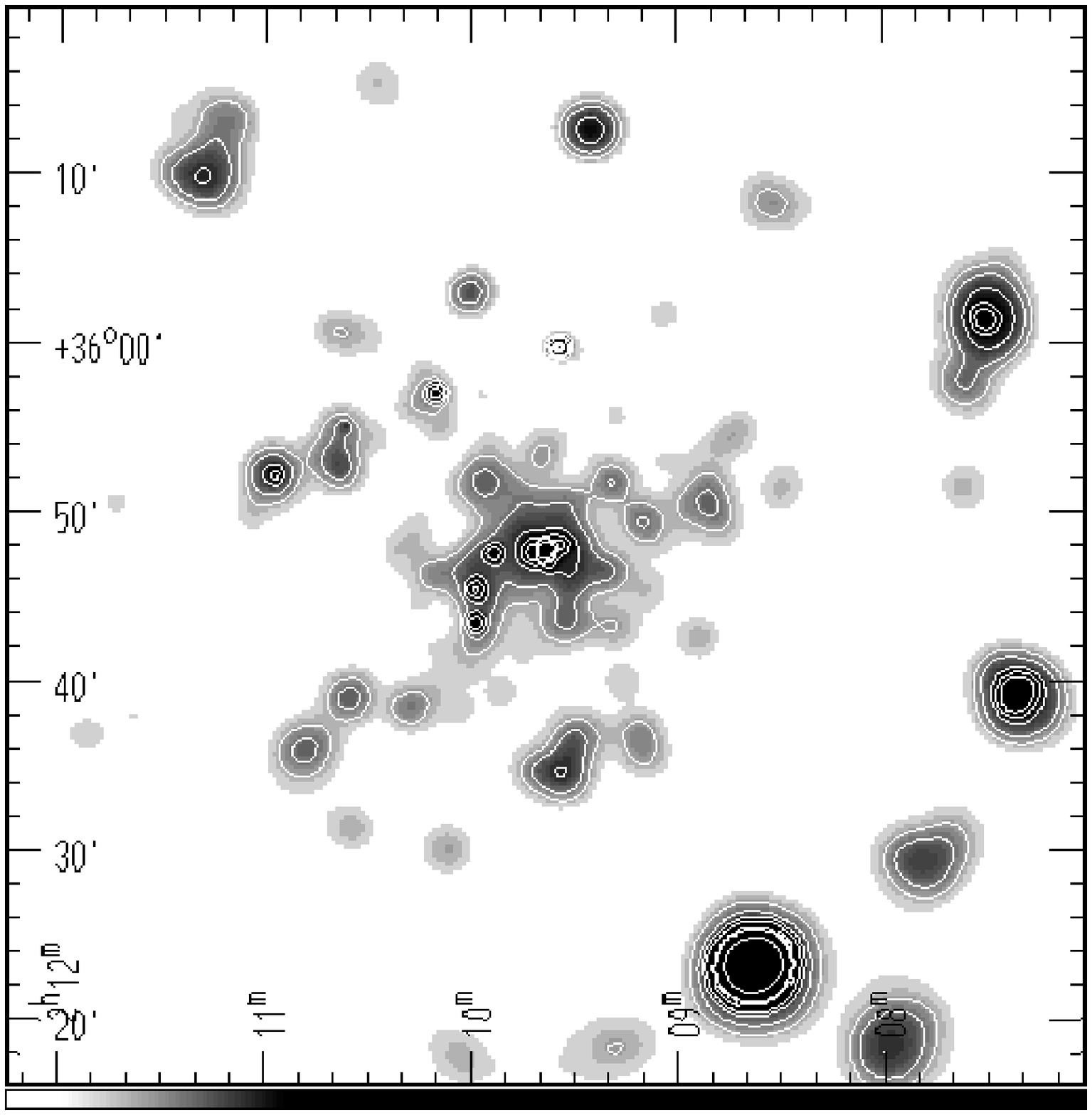}
}
\centerline{\epsfxsize=.5\hsize
\epsfbox{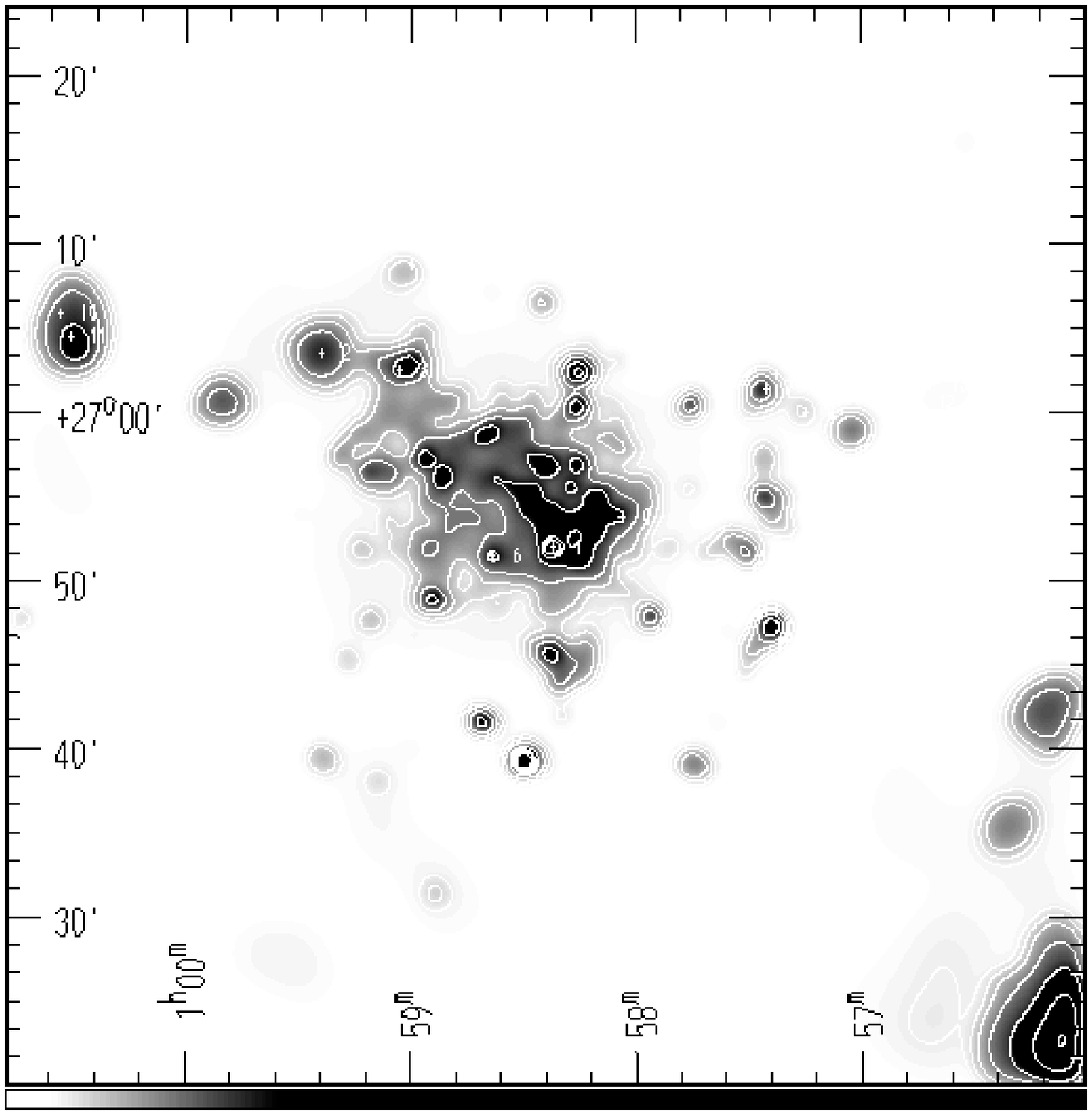}
 \hfil
\epsfxsize=.5\hsize
 \epsfbox{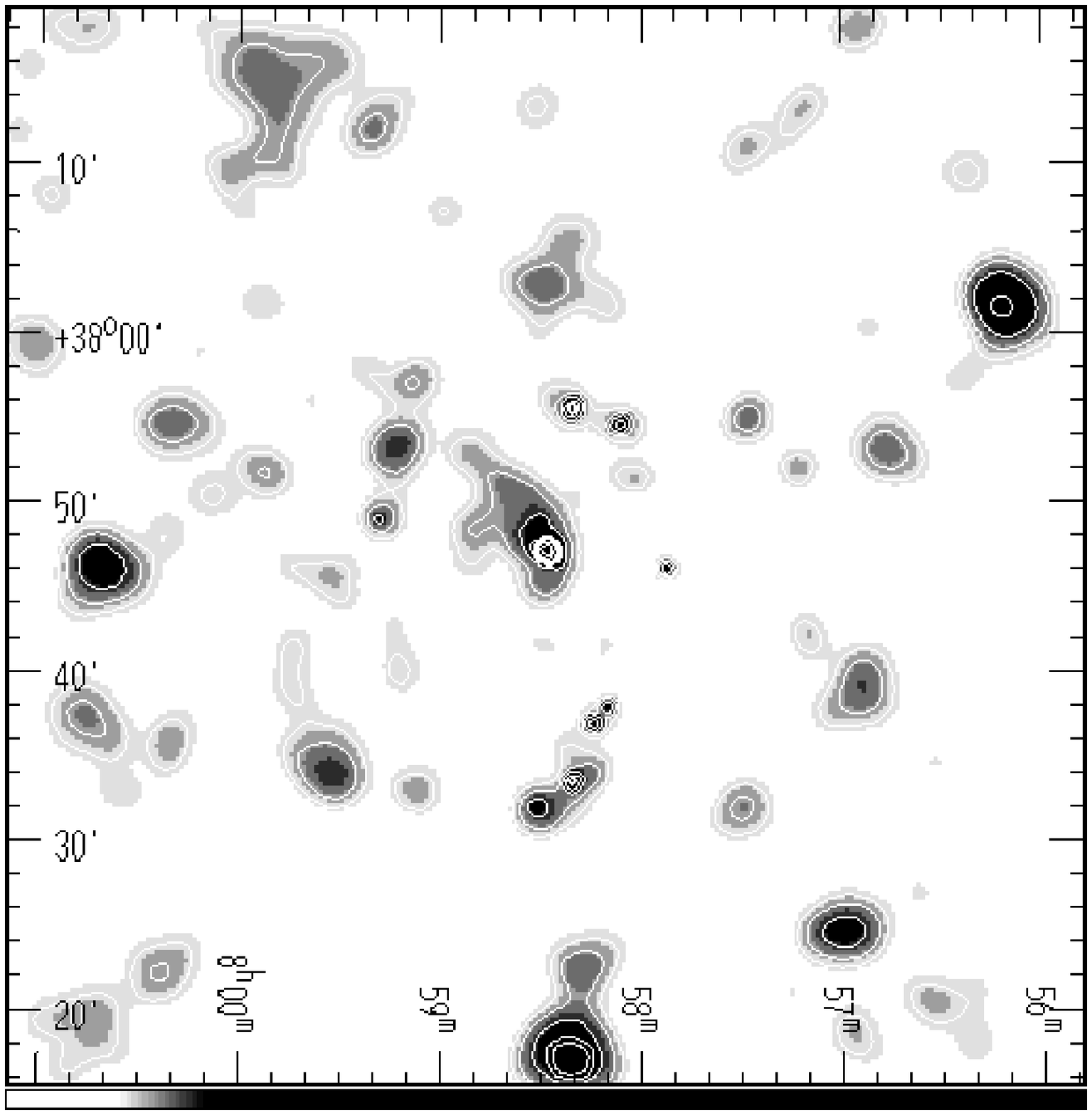}
}
\vskip -\baselineskip
{\figure{2}{ROSAT PSPC observations of 1.3 square degree fields
centered on four B2 sample radio galaxies of similar redshift, 
illustrating diversity in the amount and extent of resolved (thermal) 
X-ray emission relative to the unresolved core.
Clockwise from top left: (a) B2 1040+30, $z = 0.036$ (b) 4C 35.03, $z =
0.0375$ (c) NGC 2484, $z = 0.0413$ (d) NGC~326, $z = 0.047$.
10 arcmin corresponds to between 600  \&  770 kpc, depending on $z$, for
$H_o = 50 $~km s$^{-1}$ Mpc$^{-1}$.  [From Worrall et al.~1997, in
preparation.]
}}
}
\endfig

The importance of component separation is illustrated in Fig.~2, which
shows images from ROSAT PSPC $\simeq 20$~ks pointed exposures of four
B2 radio galaxies. Analysis of these and other sample members finds
some amount of unresolved X-ray emission in them all, but the fraction
of the total emission that is unresolved varies from a few per cent to
more than 70\%. The resolved emission varies in scale 
size from galaxy to group to cluster dimension and is associated
with hot gas.  The{\hfilneg\ \par}

\begfig 0cm
\vbox{
\vskip -\baselineskip
\centerline{\epsfxsize=.5\hsize
\epsfbox{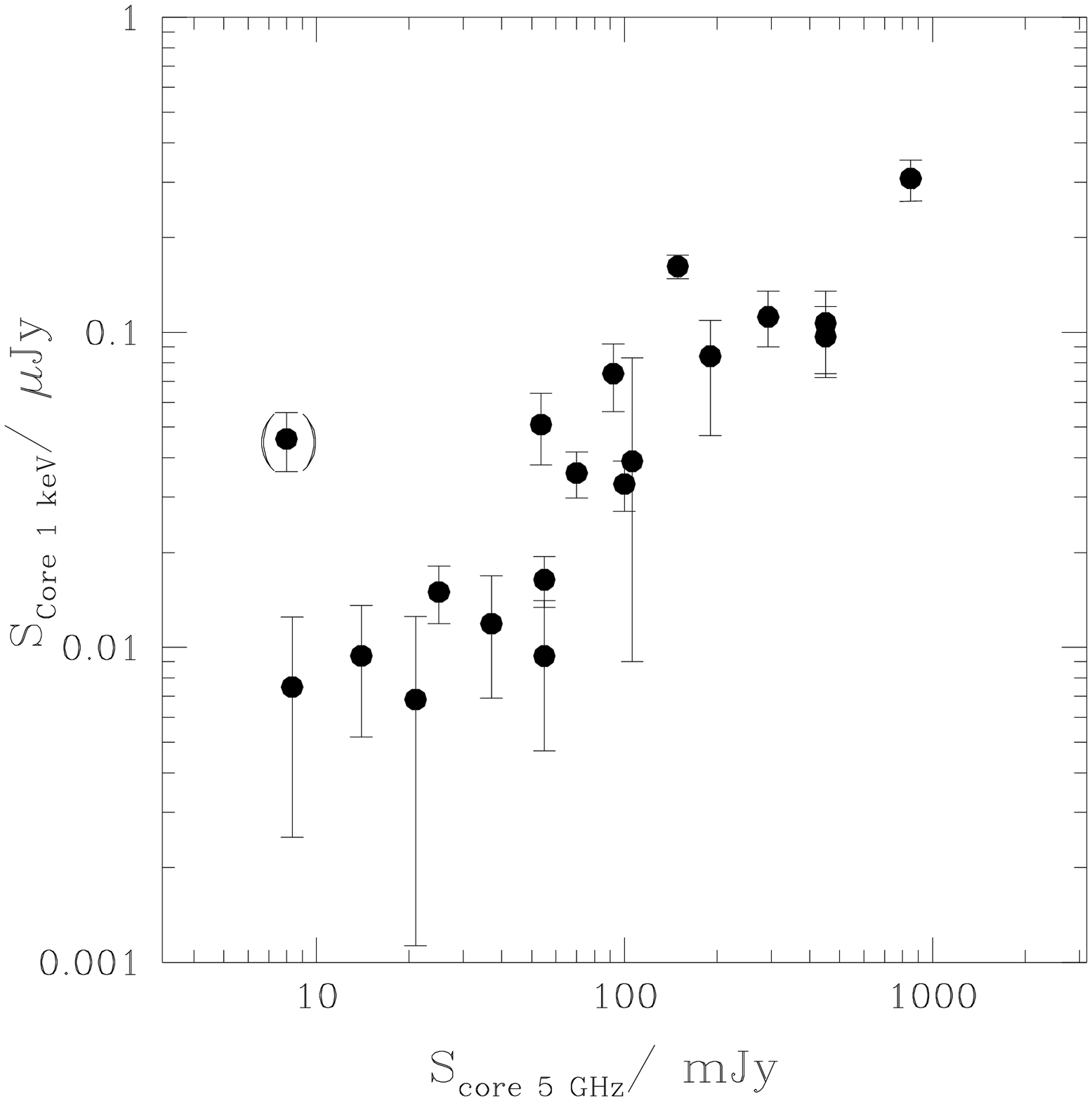}
 \hfil
\epsfxsize=.5\hsize
 \epsfbox{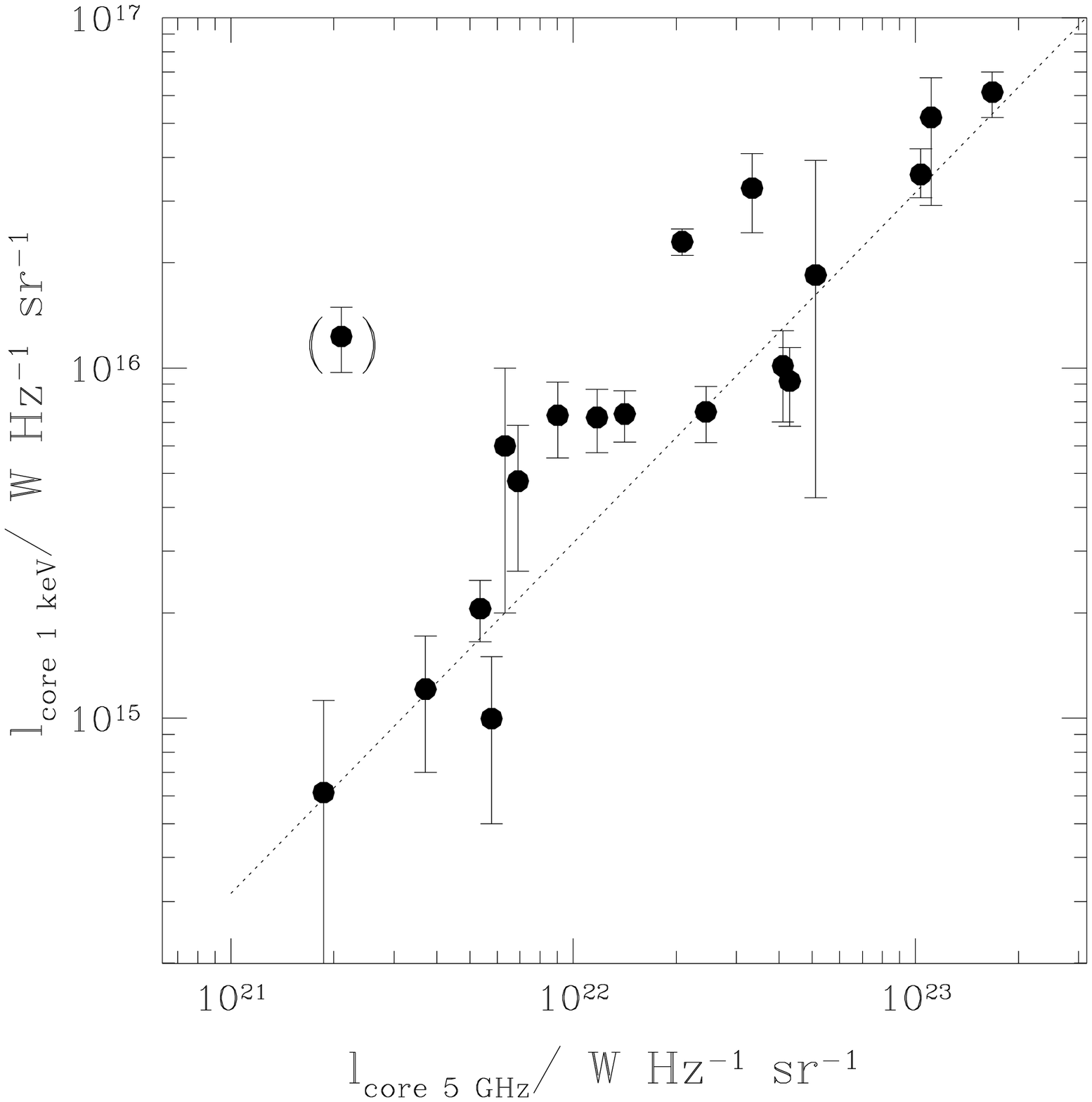}
}
\vskip -\baselineskip
{\figure{3}{
Unresolved X-ray versus core radio emission for a subset of 17 B2
sources together with NGC~6251 and NGC~4261: Flux-density plot on
the left; Luminosity density on the right [$H_o = 50 $~km s$^{-1}$
Mpc$^{-1}$] with line of slope unity
drawn for reference and giving an extrapolated radio to X-ray
spectral index $\alpha_{rx} \simeq 0.85$.  The bracketed point
represents a source which
shows an atypical X-ray structure where only part of the apparent core
X-ray flux is believed to be associated with the radio core.
}}
}
\endfig

\noindent
unresolved X-ray emission is correlated with core radio emission;
Fig.~3 extends the work of Worrall \& Birkinshaw (1994).  The earlier
claim of an X-ray/core-radio correlation from {\it Einstein\ } data
(Fabbiano et al. 1984) is now greatly strengthened by a larger sample
of X-ray detected sources and careful component separation.
The extrapolated radio to soft X-ray spectral index, $\alpha_{rx}$, 
has the same value of $\simeq 0.85$ for high-power and low-power jets
within errors.

The $\sim 5''$ angular resolution of ROSAT with the HRI is
unfortunately not met due to smearing caused by errors in ROSAT's
attitude determination.  If
the apparent resolution of some cores on scale sizes between
about $5''$ and $8''$ is taken as real, and the X-ray emission is
modelled as X-ray emitting gas in hydrostatic equilibrium, 
typical cooling times for the gas are  $< 8 \times 10^8$ yrs.
The fact that this is much less than a Hubble time ($\sim 2 \times
10^{10}$~yrs) does not favour a thermal explanation for the
small-scale X-ray emission.

Since $8''$ corresponds typically to about 8~kpc for the B2 sample,
Fig.~3 can be interpreted as the observation of jet-related X-ray
emission out to this jet radius.  However, since in low-power sources
it is not certain that the centre is obscured (see above) it could be
argued that the X-rays might come from very central regions.  To
investigate this further it is interesting to consider the radio
galaxy M87, 16~Mpc away, from which X-rays are detected in knot A at
about 0.95~kpc from the nucleus.  At a 5~GHz flux density of 1.7~Jy
and a 1~keV flux density of $0.34~\mu$Jy (Biretta et al.~1991), knot A
fits rather well on the correlation of Fig.~3.  Similarly X-ray
emission is associated with several radio features in the jet of
Cen~A, 3.5~Mpc away.  Knot~B at 1~kpc from the nucleus is perhaps the
best separated from surrounding emission in both the X-ray and radio;
Clarke et al.~(1992) report a 5~GHz flux density of 314~mJy, and the
X-ray luminosity reported by D\"obereiner et al.~(1996) corresponds to
$0.12~\mu$Jy, again fitting the correlation.  Were M87 and Cen~A at
the distances of the B2 galaxies, the knot emission would be within
the region regarded as the radio and X-ray core.  So, in at least two
nearby cases jet-related X-ray emission is present at the expected
level.  Birkinshaw \& Worrall (1993) point out that the X-ray
non-detection of a radio feature at $\sim 10$~kpc from the core of the
more distant $z = 0.024$ radio galaxy NGC~6251 is at a jet radius
where local electron acceleration is not required.  This X-ray
non-detection, which is inconsistent with Fig.~3, may then be
indicative of the changing conditions at larger jet radii which result
in the B2 sample displaying a poor correlation of {\it total\ } radio
power and core X-ray emission.

Results show measurable levels of jet-related X-ray emission in
low-power radio galaxies, with $\alpha_{rx} \simeq
0.85$. Radio-selected BL Lac objects from the 1~Jy sample have a
similar $\bar \alpha_{rx}$ (Sambruna et al.~1996) and fit on an
extrapolation of Fig.~3 but with a dispersion in $S_x$ of a factor $>
50$ whose size, compared to the tight correlation for quasars
(Fig.~1), needs accounting for when different emission models are
applied to the data.

\vskip -0.5\baselineskip\titled{Acknowledgements.}
I thank Mark Birkinshaw, Tino Canosa, and Martin Hardcastle
for discussions and for contributing to results reported in this
paper.  I thank the organizers for a stimulating workshop and am
grateful for support from PPARC grant GR/K98582 \& NASA grant NAG
5-1882.
\vskip -0.5\baselineskip
\begrefchapter{References}
\ref Arnaud, K.A., Johnstone, R.M., Fabian, A.C. et al.~1987, 
 MNRAS, 227, 241
\ref Biretta, J.A., Stern, C.P. \& Harris, D.E.~1991, AJ, 101, 1632
\ref Birkinshaw, M. \& Worrall, D.M.~1993, ApJ, 412, 568
\ref Browne, I.W.A. 1989, in `BL~Lac Objects',
  (Springer-Verlag), 401
\ref Carilli, C.L. \& Barthel, P.D.~1996, A\&A Rev.~7, 1
\ref Clarke, D.A., Burns, J.O. \& Norman, M.L.~1992, ApJ, 395, 444
\ref Crawford, C.S. \& Fabian A.C.~1996, MNRAS, 282, 1483
\ref D\"obereiner, S., Junkes, N., Wagner, S.J. et al.~1996, ApJL, 470, L15
\ref Fabbiano, G., Miller, L., Trinchieri, G., Longair, M.~\& Elvis,
  M. 1984, ApJ, 277, 115
\ref Harris, D.E., Perley, R.A. \& Carilli, C.L.~1994, IAU Symp.~159
  (Kluwer), 375
\ref Laing, R.A., Riley, J.M. \& Longair, M.S.~1983, MNRAS, 204, 151
\ref Padovani, P.~\& Urry, C.M. 1990, ApJ, 356, 75
\ref Sambruna, R.M., Maraschi, L.~\& Urry, C.M.~1996, ApJ, 463, 444
\ref Ueno, S., Koyama, K., Nishida, M., Yamauchi, S. \& Ward, M.~1994,
   ApJ, 431, L1
\ref Ulrich, M.-H.~1989, in `BL Lac Objects' (Springer-Verlag), 45
\ref Wilkes, B.J. \& Elvis, M.~1987, ApJ, 323, 243
\ref Worrall, D.M. \& Birkinshaw, M. 1994, ApJ, 427, 134
\ref Worrall, D.M., Giommi, P., Tananbaum, H. \& Zamorani, G.~1987,
  ApJ, 313, 596
\ref Worrall, D.M., Lawrence, C.R., Pearson, T.J. \& Readhead,
  A.C.S.~1994, ApJL, 420, L17
\ref Zamorani, G.~1984, in `VLBI \& Compact Radio Sources', 
  IAU Symp.~110 (Reidel), 85
\endref

\byebye